# Absolute Properties of the Eclipsing Binary Star V501 Herculis

Claud H. Sandberg Lacy[1] and Francis C. Fekel[2,3]

[1]Physics Department, University of Arkansas, Fayetteville, AR 72701, USA;

clacy@uark.edu

[2]Center of Excellence in Information Systems, Tennessee State University, Nashville, TN

37209, USA; fekel@evans.tsuniv.edu

# ABSTRACT

V501 Her is a well detached G3 eclipsing binary star with a period of 8.597687 days for which we have determined very accurate light and radial-velocity curves by using robotic telescopes. Results of these data indicate that the component stars have masses of $1.269 \pm 0.004$ and $1.211 \pm 0.003$ solar masses, radii of $2.001 \pm 0.003$ and $1.511 \pm 0.003$ solar radii, and temperatures of $5683 \pm 100$K and $5720 \pm 100$K, respectively. Comparison with the Yonsei-Yale series of evolutionary models results in good agreement at an age of about 5.1 Gyr for a somewhat metal rich composition. Those models indicate that the more massive, larger, slightly cooler star is just beyond core hydrogen exhaustion while the less massive, smaller, slightly hotter star has not quite reached core hydrogen exhaustion. The orbit is not yet circularized, and the components are rotating at or near their pseudosynchronous velocities. The distance to the system is $420 \pm 30$ pc.

Key Words: binaries: eclipsing – binaries: spectroscopic – stars: fundamental parameters

– stars: individual (V501 Her)

1.  Introduction

The discovery of the star we now know as V501 Her was announced by Kippenhahn (1955) who gave the range of variation of the new variable star (S4154 Her) as 0.5 mag, and provided a finder chart for it.  Gessner (1966) gave the magnitude range as 10.5–11.0.  The first accurate eclipse ephemeris was due to Kreiner (2004).  There have been a number of accurate times of minima published, which allow us to determine a better eclipse ephemeris (Sec. 2) for use in analyzing the light curves from our robotic telescopes, the URSA and NFO WebScopes (Sec. 2).  We have measured radial velocities of the system, mainly by using a robotic telescope at Fairborn Observatory, and give the analysis in Sec. 3.  The absolute properties are presented in Sec. 4, along with their interpretation according to current theory, where we find good agreement with the observations.

2.  Photometric data and analysis

Accurate photographic and CCD times of minima were gathered from the literature and used to determine an eclipse ephemeris (visual estimates and a couple of photoelectric estimates were not used because of their much larger observational errors and systematic deviations).  These times and the residuals from the fits are given in Table 1.  The least-squares best fit eclipse ephemeris is:

HJD Min I = 2,455,648.59428(20) + 8.5976870(10) E             (1)

where the uncertainty in the last digits is given in parentheses. Relative to this ephemeris, secondary eclipse occurs at a phase of 0.4791(2), so the orbit is eccentric.

Table 1. Dates of Minima for V501 Her. Type 1 eclipses are the deeper eclipses. Cycle numbers are relative to the primary eclipse ephemeris (Eq. 1).

| HJD−2400000 | Eclipse Type | Uncertainty (days) | O−C (days) | Cycle Number | Reference |
|---|---|---|---|---|---|
| 29752.393 | 1 | 0.050 | 0.032 | −3012 | 1 |
| 30431.540 | 1 | 0.050 | −0.038 | −2933 | 1 |
| 30990.419 | 1 | 0.050 | −0.009 | −2868 | 1 |
| 31652.464 | 1 | 0.050 | 0.014 | −2791 | 1 |
| 53511.8879 | 2 | 0.0003 | −0.0011 | −248.5 | 2 |
| 53550.7611 | 1 | 0.0010 | 0.0025 | −244 | 2 |
| 53963.4443 | 1 | 0.0013 | −0.0033 | −196 | 3 |
| 55278.8917 | 1 | 0.0009 | −0.0020 | −43 | 4 |
| 55321.8833 | 1 | 0.0005 | 0.0011 | −38 | 4 |
| 55351.7943 | 2 | 0.0007 | 0.0002 | −34.5 | 4 |
| 55364.8705 | 1 | 0.0005 | −0.0001 | −33 | 4 |
| 55648.59428 | 1 | 0.00087 | 0.00000 | 0 | 5 |
| 55712.8993 | 2 | 0.0006 | 0.0024 | 7.5 | 6 |
| 56069.8817 | 1 | 0.0011 | 0.0008 | 49 | 7 |
| 56099.7924 | 2 | 0.0003 | −0.0004 | 52.5 | 7 |
| 56460.8970 | 2 | 0.0005 | 0.0013 | 94.5 | 8 |

**References.** (1) Gessner 1966; (2) Krajci 2006; (3) Hubscher & Walter 2007; (4) Lacy 2011; (5) Zasche et al. 2011; (6) Lacy 2012; (7) Lacy 2013; (8) Lacy 2014.

Differential photometry in the V filter was performed with the URSA and NFO WebScopes, as described by Lacy, Torres, & Claret (2012) and Grauer, Neely, & Lacy (2008). The comparison stars used were TYC 2606-0980-1 (V=9.54 mag) and TYC 2606-0827-1 (V=10.67 mag). The sum of the comparison star fluxes was used as the reference flux, and this was called comps, so the differential magnitude ΔV was var-comps. Each night the differences in magnitudes between the comparison stars was measured, and these differences had a standard deviation of 0.011 mag (URSA) and 0.007 mag (NFO).  We expected that these values would be close to the standard error found after light curve modelling of the eclipsing binary (see Table 4), which they are.  A total of 2939 magnitudes was obtained from the URSA WebScope from 2010 February 18 to 2012 July 18, and a total of 7267 magnitudes was obtained from the NFO WebScope from 2009 April 24 to 2014 April 1.  The differential photometry is given in Tables 2 & 3.

Table 2.  URSA Differential photometry of V501 Her

| HJD-2400000 | ΔV (mag) |
|---|---|
|  |  |
| 55245.95343 | 1.953 |
| 55245.95531 | 1.940 |
| 55245.95719 | 1.946 |
| 55245.95907 | 1.933 |
| 55245.96096 | 1.942 |

(This table is available in its entirety in machine-readable and Virtual Observatory (VO) forms in the online journal. A portion is shown here for guidance regarding its form and content.)

Table 3.  NFO Differential Photometry of V501 Her

| HJD-2400000 | ΔV (mag) |
|---|---|
|  |  |
| 54945.75766 | 1.911 |
| 54945.75984 | 1.901 |
| 54945.76202 | 1.892 |
| 54945.76420 | 1.904 |
| 54945.76643 | 1.899 |

(This table is available in its entirety in machine-readable and Virtual Observatory (VO) forms in the online journal. A portion is shown here for guidance regarding its form and content.)

Nightly photometric zero-point corrections on the order of 0.01 mag were determined from the data based on preliminary orbital fits, then the data were corrected for this instrumental effect.  These variations are due to imprecise centering from night to night and variations in responsivity across the field of view.  The final light curves and the residuals to the fits are shown in Fig. 1-3.

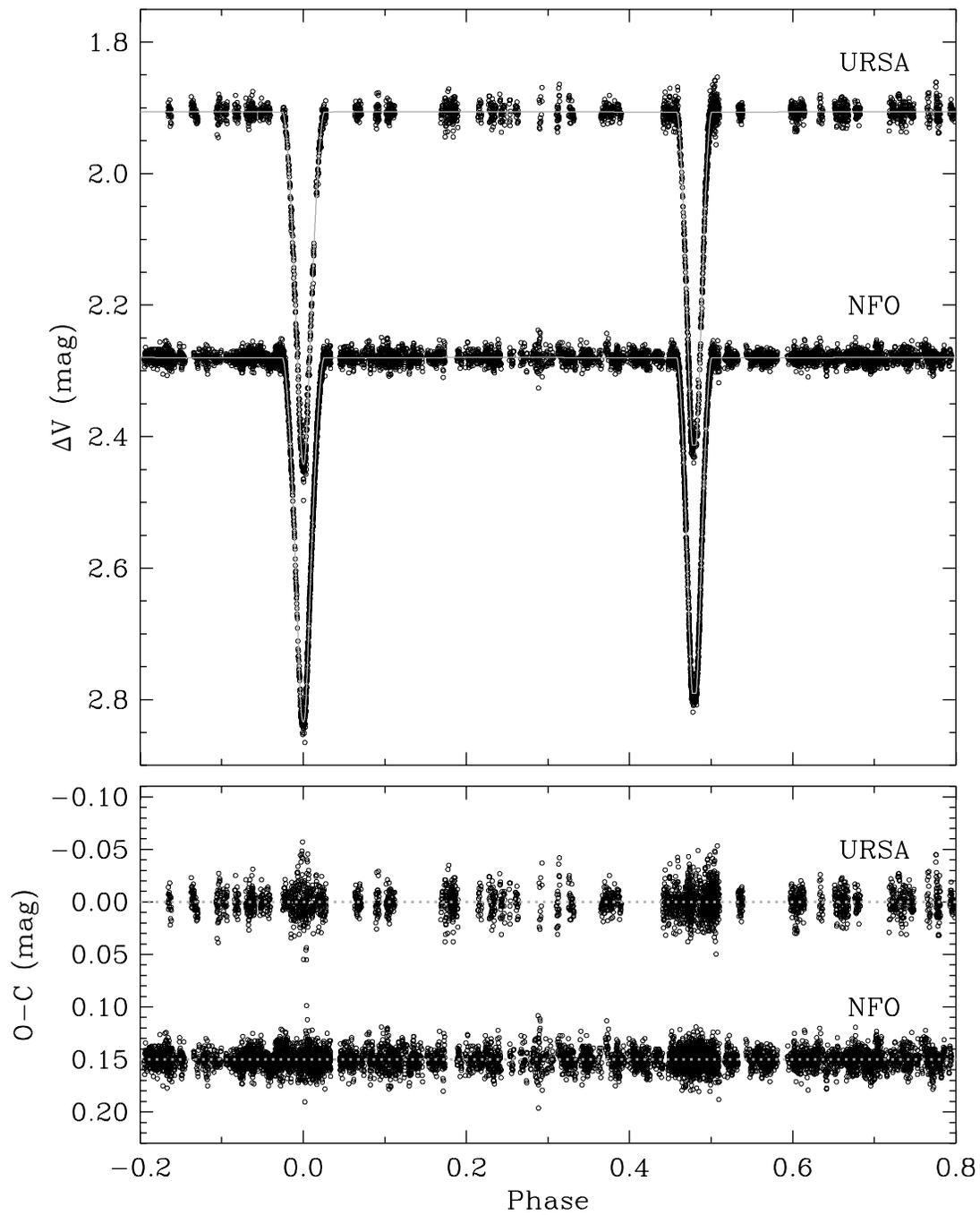

Figure 1. Upper panel, V-band light curves of V501 Her from URSA and NFO WebScope images. The observational points for the NFO telescope have been offset

for clarity. The fitted photometric model is shown in grey. Lower panel, residuals from the fits to the URSA and NFO data. Again, the NFO points are offset for clarity.

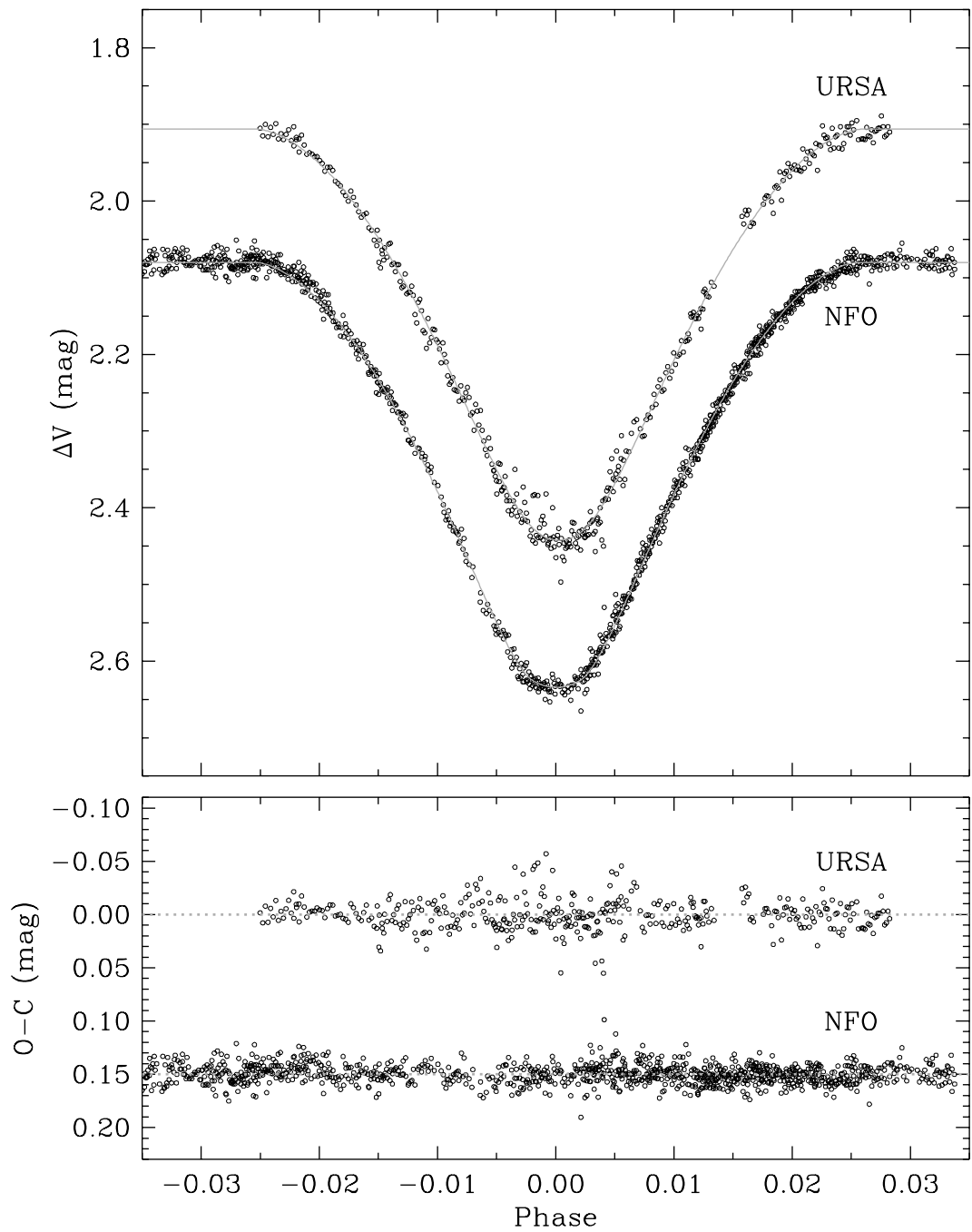

Figure 2. V-band observations in the region around the annular primary eclipse of V501 Her.

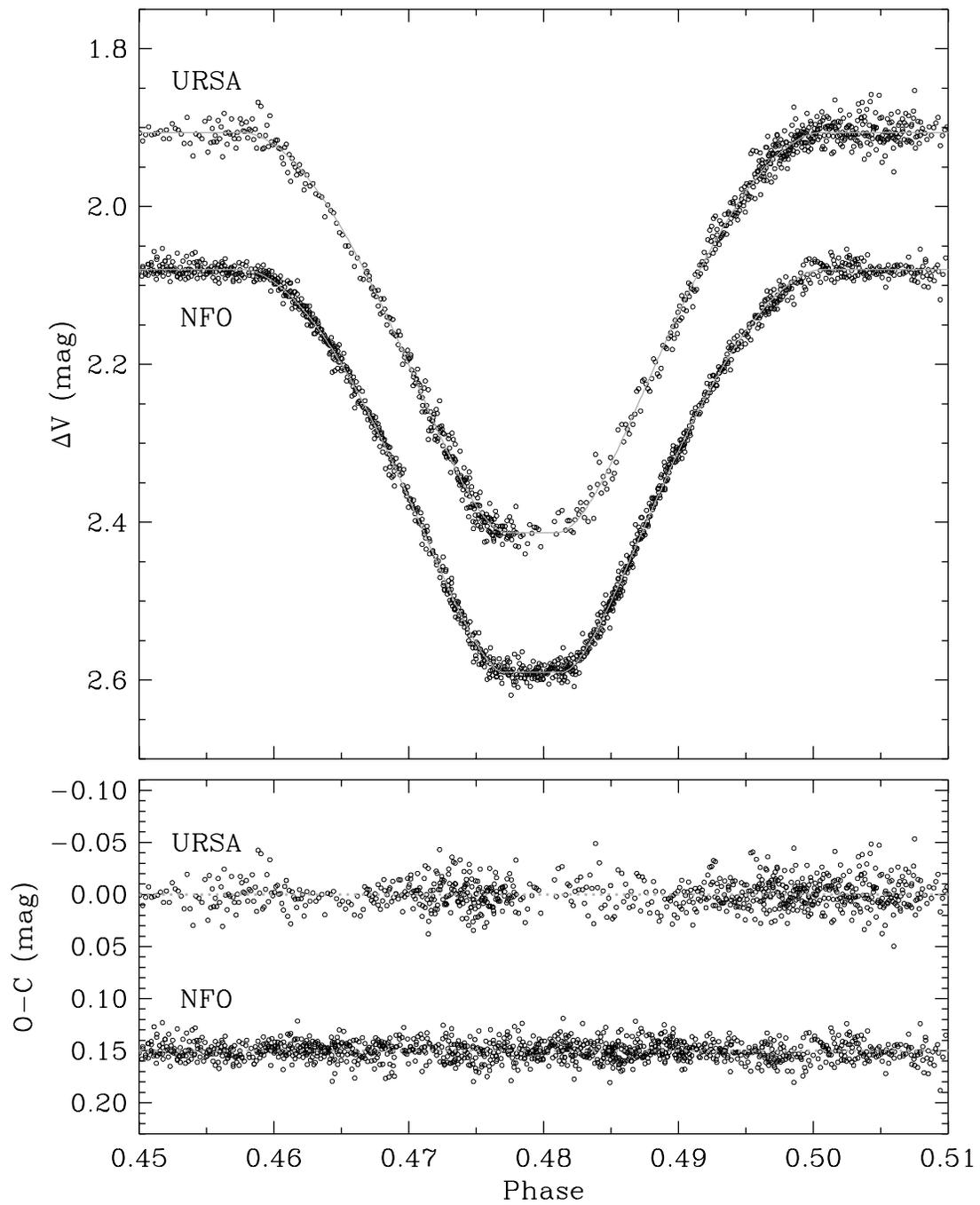

Figure 3. V-band light curve of V501 Her near the total secondary eclipse.

The light curve model fitted to the photometric data is the Nelson-Davis-Etzel (NDE) model (Nelson & Davis 1972, Popper & Etzel 1981, Southworth, Maxted, & Smalley 2004, North & Zahn 2004) as implemented in the code *jktebop*, written by John Southworth. The main variables of the model are $J_B$, the unitless central surface brightness of the smaller star (star B) relative to that of the larger star (star A); $r_A+r_B$, the unitless sum of radii of the stars relative to the semi-major axis of the orbit; k, the unitless ratio of radii, $r_B/r_A$; $u_A$ and $u_B$, the unitless linear limb-darkening coefficients; i, the orbital inclination in degrees; q, the the mass ratio ($m_B/m_A = 0.9546$) is adopted from the spectroscopic orbit, and not varied; $L_A$ and $L_B$, the unitless V-band fluxes of the components relative to the V-band flux from the binary system at quadrature phase; $\sigma$, the standard error of the residuals from the fitted orbit, in milli-magnitudes; N, the number of observations used in the fit; and the number of Corrections applied to compensate for the nightly shifts in the magnitude zero-point. The auxiliary unitless values of gravity darkening exponents $\beta_1$ (0.37, both components) were taken from theory (Claret 1998) based on the components' temperatures. The mass ratio is adopted from the spectropic orbit (0.9546). There was no detectable third-light in the light curve fit, so we have assumed that the only light source is the binary star. Non-linear limb-darkening coefficients of the log type (Claret 2000) were tried instead of the linear limb-darkening coefficients, but the fitted values were not significantly different and the residual variance was worse for the URSA data and very slightly better (0.1 %) for the NFO data set. The fitted parameters are shown in Table 4. In this binary system, star A, the stronger-lined, more massive, larger, cooler, faster rotating star is the one being eclipsed at primary eclipse. This situation is made possible by the eccentricity and orientation of the orbit. It is worth pointing out that secondary eclipse is a total eclipse in this binary. The annular primary eclipse is shown in Figure 2 and the total secondary eclipse is shown in Figure 3. No spectroscopic observations were made in secondary eclipse, however.

Table 4. Fitted orbital parameters of the NDE model for V501 Her.

| Parameters | URSA | NFO | Adopted |
|---|---|---|---|
| $J_B$ | 1.0724 ± 0.0037 | 1.0412 ± 0.0016 | 1.046 ± 0.006 |
| $r_A+r_B$ | 0.14771 ± 0.00047 | 0.14686 ± 0.00019 | 0.14698 ± 0.00019 |
| k | 0.7427 ± 0.0024 | 0.7570 ± 0.0010 | 0.7550 ± 0.0020 |
| i (degrees) | 88.892 ± 0.054 | 89.144 ± 0.027 | 89.11 ± 0.04 |
| $u_A = u_B$ | 0.728 ± 0.032 | 0.629 ± 0.014 | 0.64 ± 0.04 |
| $r_A$ | 0.08476 ± 0.00029 | 0.08358 ± 0.00012 | 0.08375 ± 0.00011 |
| $r_B$ | 0.06295 ± 0.00023 | 0.06328 ± 0.00010 | 0.06323 ± 0.00010 |
| $L_B/L_A$ | 0.591 ± 0.010 | 0.5966 ± 0.0045 | 0.595 ± 0.004 |
| $L_A$ | 0.6279 ± 0.0043 | 0.6258 ± 0.0018 | 0.6266 ± 0.0016 |
| $L_B$ | 0.3713 ± 0.0030 | 0.3734 ± 0.0013 | 0.3725 ± 0.0015 |
| σ (mmag) | 12.3 | 8.7 | 7.5 |
| N | 2939 | 7267 | |
| Corrections | 70 | 228 | |

* Note: The orbital period and epoch have been fixed at the photometric values in equation (1): HJD Min I = 2,455,648.59428(20) + 8.5976870(10) E.

The adopted values in Table 4 were determined by weighted averaging, conservatively increased by adding in quadrature half the difference between URSA and NFO values.

3.  Spectroscopic observations and analysis

From 2010 March through 2014 April we obtained 63 spectrograms of V501 Her at two different observatories. The vast majority of our spectra, 58 of them, were acquired at Fairborn Observatory in southeast Arizona with the Tennessee State University 2 m Automatic Spectroscopic Telescope (AST) and a fiber-fed echelle spectrograph (Eaton & Williamson 2007). Through 2011 June we used a 2048 x 4096 SITe ST-002A CCD with 15 μm pixels as the detector. Eaton & Williamson (2007) have explained the reduction of the raw spectra and the spectrum wavelength calibration. Those echelle spectrograms have 21 orders that cover the wavelength range 4920–7100 Å. About two-thirds of the spectra have a resolution of 0.4 Å, while the rest have a somewhat higher resolution of 0.24 Å. The signal-to-noise ratios of those spectra ranged from 20 to 30.

During the summer of 2011, the SITe CCD detector and dewar of the 2 m AST were replaced by a Fairchild 486 CCD, which consists of an array of 4096 x 4096 15 μm pixels, and a new dewar. Fekel et al. (2013) have more extensively discussed the AST improvements. The spectrograms that were obtained with the new CCD have 48 orders that cover wavelengths that range from 3800--8260 Å. All our new spectra have a resolution of 0.4 Å with signal-to-noise ratios of about 35–50.

In 2010 we acquired an additional five spectra with the Kitt Peak National Observatory

(KPNO) coude feed telescope, coude spectrograph, and a TI CCD. Those spectrograms were centered at 6430 Å, have a wavelength range of 84 Å, and a resolution of 0.21 Å or a resolving power of over 30,000. The signal-to-noise ratios are about 75. A sample spectrum is shown in Figure 4.

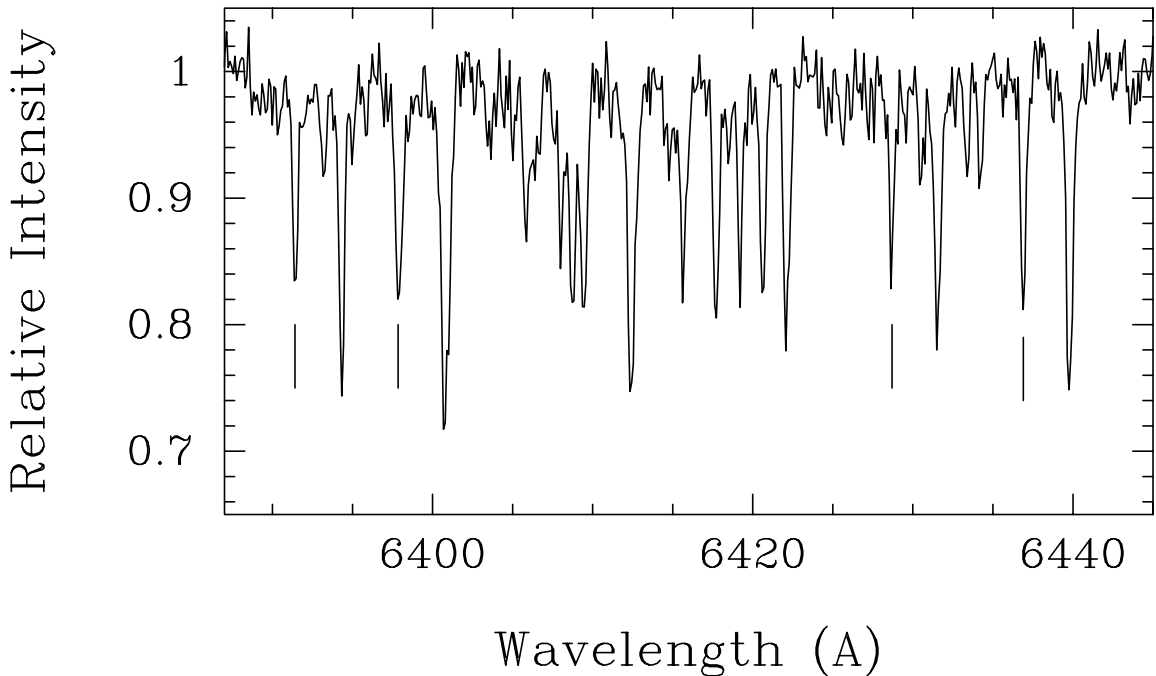

Figure 4 - Spectrum of V501 Her in the 6430 Å region taken at KPNO. Tick marks indicate several blueshifted lines of component B, the less massive and smaller star.

Fekel et al. (2009) have given a general explanation of the velocity measurement of our AST echelle spectra. For V501 Her the solar-type star line list, which consists of 168 moderate strength lines that range in wavelength from 4966 A to 6768 A, was used. We determined our velocities by fitting the individual lines with rotational broadening functions (Lacy & Fekel 2011) and allowed both the depth and width of the line fits to vary. Measurements of several IAU solar-type velocity standards show that our AST velocities have a zero-point offset of -0.6 km/s compared to the results of Scarfe (2010).

Thus, we have added 0.6 km/s to each Fairborn velocity of V501 Her. The Fairborn radial velocity observations are listed in Table 5.

Using Gaussian functions to fit the profiles, we measured our KPNO velocities by cross correlation with respect to IAU radial velocity standard stars HR 5694 or HR 7560. Their velocities of 54.4 and 0.0 km/s were adopted from Scarfe (2010). The KPNO radial velocity observations are also given in Table 5.

Table 5. Radial velocity observations and residuals from the fitted orbit.

| HJD-2,400,000 | Phase from Periastron | $RV_A$ (km/s) | $(O-C)_A$ (km/s) | $RV_B$ (km/s) | $(O-C)_B$ (km/s) | Observatory |
|---|---|---|---|---|---|---|
| 55259.976 | 0.366 | 23.0 | 0.4 | -97.1 | -0.4 | Fairborn |
| 55299.859 | 0.005 | -58.9 | 0.2 | -9.8 | 1.4 | Fairborn |
| 55300.900 | 0.126 | -0.3 | -0.3 | -73.2 | -0.1 | Fairborn |
| 55314.957 | 0.761 | -100.3 | -0.7 | 30.9 | -0.4 | KPNO |
| 55323.880 | 0.799 | -104.3 | 0.9 | 37.8 | 0.7 | Fairborn |
| 55339.927 | 0.666 | -72.7 | 0.7 | 5.0 | 1.2 | Fairborn |
| 55355.789 | 0.511 | -17.1 | 0.6 | -54.2 | 0.4 | Fairborn |
| 55366.873 | 0.800 | -105.2 | 0.0 | 36.9 | -0.2 | KPNO |
| 55367.851 | 0.913 | -96.0 | -0.1 | 26.9 | -0.4 | KPNO |
| 55369.728 | 0.132 | 1.4 | -0.8 | -76.4 | -1.0 | Fairborn |
| 55369.802 | 0.140 | 5.1 | -0.4 | -78.5 | 0.4 | KPNO |
| 55370.858 | 0.263 | 30.6 | -0.3 | -104.7 | 0.7 | KPNO |
| 55638.920 | 0.442 | 4.5 | 0.0 | -78.2 | -0.4 | Fairborn |
| 55666.898 | 0.696 | -84.5 | -1.4 | 13.8 | -0.1 | Fairborn |
| 55683.816 | 0.663 | -72.0 | 0.7 | 2.9 | -0.1 | Fairborn |
| 55710.782 | 0.800 | -105.2 | 0.1 | 38.0 | 0.8 | Fairborn |
| 55984.907 | 0.683 | -79.3 | -0.1 | 10.1 | 0.2 | Fairborn |
| 55991.986 | 0.507 | -16.6 | -0.2 | -56.0 | -0.1 | Fairborn |
| 56001.906 | 0.661 | -71.7 | 0.0 | 2.5 | 0.5 | Fairborn |
| 56022.840 | 0.095 | -13.9 | -0.3 | -60.1 | -1.2 | Fairborn |
| 56035.850 | 0.609 | -53.8 | -0.5 | -17.0 | 0.3 | Fairborn |
| 56036.839 | 0.724 | -91.4 | -0.4 | 21.3 | -0.9 | Fairborn |
| 56038.874 | 0.960 | -79.6 | 0.1 | 10.4 | 0.0 | Fairborn |
| 56039.874 | 0.077 | -22.9 | -0.3 | -49.3 | 0.1 | Fairborn |
| 56040.853 | 0.191 | 20.6 | -0.3 | -95.6 | -0.6 | Fairborn |
| 56042.874 | 0.426 | 9.4 | 0.4 | -82.3 | 0.2 | Fairborn |
| 56046.975 | 0.903 | -98.6 | 0.0 | 30.4 | 0.2 | Fairborn |
| 56051.972 | 0.484 | -8.0 | 0.6 | -64.1 | -0.0 | Fairborn |
| 56053.782 | 0.694 | -82.6 | 0.0 | 12.9 | -0.6 | Fairborn |
| 56054.761 | 0.808 | -106.1 | -0.1 | 38.1 | 0.2 | Fairborn |
| 56055.766 | 0.925 | -92.4 | 0.1 | 23.3 | -0.5 | Fairborn |
| 56058.751 | 0.272 | 31.0 | -0.1 | -104.8 | 0.9 | Fairborn |
| 56059.941 | 0.411 | 12.5 | -0.4 | -86.5 | 0.1 | Fairborn |
| 56060.853 | 0.517 | -20.0 | -0.2 | -51.9 | 0.4 | Fairborn |
| 56061.736 | 0.619 | -57.2 | 0.0 | -13.0 | 0.1 | Fairborn |
| 56062.739 | 0.736 | -94.5 | -0.4 | 25.6 | 0.1 | Fairborn |
| 56063.740 | 0.853 | -106.3 | -0.1 | 38.1 | -0.0 | Fairborn |
| 56064.733 | 0.968 | -76.4 | 0.0 | 6.5 | -0.5 | Fairborn |
| 56065.754 | 0.087 | -17.4 | 0.3 | -54.8 | -0.2 | Fairborn |
| 56066.740 | 0.201 | 23.7 | 0.4 | -97.7 | -0.1 | Fairborn |
| 56067.728 | 0.316 | 29.5 | 0.1 | -103.6 | 0.4 | Fairborn |
| 56068.721 | 0.432 | 7.4 | 0.2 | -80.4 | 0.3 | Fairborn |
| 56070.716 | 0.664 | -72.4 | 0.5 | 3.4 | 0.2 | Fairborn |
| 56071.771 | 0.787 | -103.6 | 0.1 | 35.9 | 0.3 | Fairborn |
| 56072.767 | 0.902 | -98.7 | -0.0 | 30.4 | 0.1 | Fairborn |
| 56073.783 | 0.021 | -52.3 | -1.0 | -18.6 | 0.8 | Fairborn |
| 56074.713 | 0.129 | 1.3 | 0.3 | -73.4 | 0.8 | Fairborn |
| 56075.750 | 0.249 | 29.8 | -0.3 | -104.6 | 0.0 | Fairborn |
| 56076.694 | 0.359 | 24.1 | 0.3 | -98.1 | -0.0 | Fairborn |
| 56088.904 | 0.779 | -102.2 | 0.5 | 34.4 | -0.1 | Fairborn |
| 56134.773 | 0.114 | -5.0 | 0.0 | -68.1 | -0.2 | Fairborn |
| 56194.714 | 0.086 | -17.7 | 0.3 | -54.2 | 0.1 | Fairborn |
| 56227.647 | 0.917 | -95.3 | -0.3 | 26.4 | -0.0 | Fairborn |
| 56316.034 | 0.197 | 22.2 | -0.2 | -96.9 | -0.3 | Fairborn |
| 56351.962 | 0.376 | 20.2 | -0.6 | -94.6 | 0.3 | Fairborn |
| 56388.836 | 0.665 | -73.2 | -0.1 | 3.1 | -0.3 | Fairborn |
| 56409.856 | 0.109 | -7.7 | -0.5 | -66.1 | -0.5 | Fairborn |
| 56432.741 | 0.771 | -102.2 | -0.8 | 32.9 | -0.2 | Fairborn |
| 56457.719 | 0.676 | -77.1 | -0.1 | 7.5 | -0.0 | Fairborn |
| 56578.666 | 0.744 | -95.5 | 0.4 | 27.3 | -0.1 | Fairborn |
| 56688.015 | 0.462 | -1.6 | 0.1 | -70.4 | 0.9 | Fairborn |
| 56710.017 | 0.021 | -50.7 | 0.3 | -20.1 | -0.4 | Fairborn |
| 56752.848 | 0.003 | -60.1 | 0.1 | -9.9 | 0.1 | Fairborn |

(This table is available in its entirety in machine-readable and Virtual Observatory (VO) forms in the online journal.

From the Fairborn Observatory spectra the average projected rotational velocities of the components are 14.6 and 12.8 km/s for stars A and B, respectively, with estimated uncertainties of 1.0 km/s. From our small number of KPNO spectra we measured $v_{rot} \sin i$ values with the procedure of Fekel (1997). A macroturbulence of 3 km/s was assumed, and velocities of 14.0 and 12.2 km/s were determined for components A and B, respectively, with estimated uncertainties of 1.0 km/s. A straight average of the two results produces projected rotational velocities of 14.3 and 12.5 km/s for components A and B, respectively. From our Fairborn Observatory spectra we determined the spectroscopic light ratio of the two components $L_B/L_A$ to be $0.58 \pm 0.02$ at 6000 Å.

Strassmeier & Fekel (1990) identified critical line ratios in the 6430 Å region for spectral type classification. Thus, KPNO spectra of a variety of late-F, G, and early-K dwarfs and subgiants, with spectral types taken from the list of Keenan & McNeil (1989), were rotationally broadened, shifted in wavelength space, and added together (Barden 1985) in an attempt to reproduce the KPNO spectrum of V501 Her in the 6430 A region. Good fits to the important line ratios occur for mid-G dwarfs. Our best spectrum addition combination is with 20 LMi = HR 3951 used for both components. That star is classified as G3Va by Keenan & McNeil (1989) and is metal rich with [Fe/H] = 0.21 according to Taylor (2005). However, the lines of V501 Her are stronger than those of 20 LMi, suggesting that V501 Her is even more metal rich. The mean temperature of the binary is estimated from the spectral type (via Popper 1980, Table 1) and its uncertainty, and the de-reddened color indices (via Casagrande et al. 2010) as $5700 \pm 100$ K. The temperature difference is estimated from the central surface brightness $J_B$ ($\Delta F_V = 0.25 \log J_B$ and hence $\Delta T$ via Popper 1980 Table 1) as 37 K, resulting in estimates of $5683 \pm 100$

K and 5720 ± 100 K.

Spectroscopic Orbital Solution

Adopting the photometric period, we obtained preliminary orbital elements of the primary with BISP (Wolfe et al. 1967), a computer program that uses the Wilsing-Russell method to determine the elements. Separate orbits of the two components were then computed with SB1 (Barker et al. 1967), a program that uses differential corrections to improve the preliminary elements. A comparison of the two solutions shows that the two center-of-mass velocities differ by just 0.1 km/s and values of the other elements in common for the two components are also in very good agreement. From the variances of the two solutions the velocities of the primary and secondary were assigned weights of 1.0 and 0.8, respectively. We then combined the weighted velocities into a double-lined solution, using a slightly modified version of SB1. Table 6 lists the resulting orbital elements and related parameters. A solution of the combined data with the period as a free parameter resulted in P = 8.597638 ± 0.000032 days, which differs by 1.5 sigma from the more precise photometric value of 8.597687 ± 0.000001 days. Our Fairborn and KPNO radial velocities are compared with the calculated velocity curves of the primary and secondary in Figure 8, where zero phase is a time of periastron. The time of conjunction with the more massive and cooler star behind, which corresponds to primary eclipse, is at phase 0.5667.

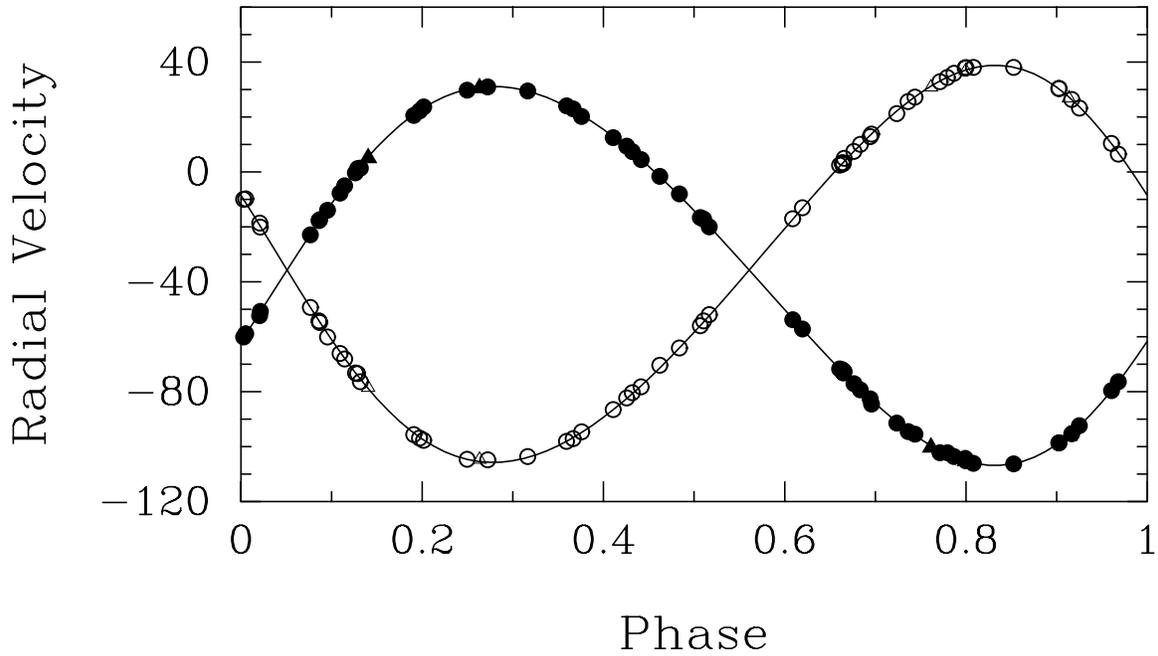

Figure 5 - Radial velocities of V501 Her compared with the computed velocity curves. Primary: filled circles (Fairborn), filled triangles (KPNO). Secondary: open circles (Fairborn), open triangles (KPNO). Zero phase is a time of periastron passage.

Table 6. Spectroscopic orbital elements.

| Parameter | Star A | Star B |
|---|---|---|
|  |  |  |
| P (days) | 8.597687 (adopted) ||
| T (HJD periastron) | 2,456,004.824 ± 0.014 ||
| e | 0.09025 ± 0.00078 ||
| ω (degrees) | 249.81 ± 0.60 ||
| K (km/s) | 68.997 ± 0.081 | 72.281 ± 0.090 |
| γ (km/s) | -35.705 ± 0.042 ||
| m sin³ i (solar masses) | 1.2696 ± 0.0032 | 1.2119 ± 0.0030 |
| a sin i (10⁶ km) | 8.124 ± 0.010 | 8.511 ± 0.011 |
| a (10⁶ km) | 16.637 ± 0.015 | |
| Mass ratio $m_B/m_A$ | 0.9546 ± 0.0034 | |
| σ (km/s) | 0.4 | 0.5 |
| Observations | 63 | 63 |

4. Absolute properties and comparison with theory

The combination of the photometric and spectroscopic analyses results in the absolute properties of Table 7.

Table 7. Absolute properties of V501 Her.

| Parameter | Star A | Star B |
|---|---|---|
| Mass (solar masses) | 1.2690 ± 0.0035 | 1.2113 ± 0.0032 |
| Radius (solar radii) | 2.0017 ± 0.0031 | 1.5112 ± 0.0027 |
| Log g (cgs) | 3.9384 ± 0.0014 | 4.1623 ± 0.0016 |
| Temperature (K) | 5683 ± 100 | 5720 ± 100 |
| Temperature difference (K) | 37 ± 8 | |
| Log L (solar units) | 0.576 ± 0.030 | 0.344 ± 0.029 |
| $F_V$* | 3.736 ± 0.007 | 3.742 ± 0.007 |
| $M_V$ (mag)* | 3.39 ± 0.07 | 3.94 ± 0.07 |
| m-M (mag)* | 8.276 ± 0.150 | |
| Distance (pc)* | 420 ± 30 | |
| Measured $v_{rot} \sin i$ (km/s) | 14.3 ± 1.0 | 12.5 ± 1.0 |
| Pseudosynchronous $v_{rot} \sin i$ (km/s) | 14.3 ± 0.1 | 10.8 ± 0.1 |

* Relies on the visual surface brightness parameter ($F_V$) calibration of Popper (1980).

The effects of interstellar dust on the distance modulus were estimated from the reddening maps of of Burstein & Heiles (1982), Schlegel et al. (1998), Drimmel et al. (2003), and Amores & Lepine (2005) as $E_{B-V}$ = 0.048 ± 0.018 mag and $A_V$ = 0.149 ±

0.056 mag. The V magnitude of the system outside eclipse was estimated from the Tycho $V_T$ and $(B-V)_T$ as $11.152 \pm 0.081$ mag. Distances derived from the magnitudes of the individual binary star components differ from the value in Table 7 by less than 4 pc.

The measured rotational velocities are probably consistent with rotational velocities that are synchronous at periastron passage ("pseudosynchronous" velocities, Giuricin et al. 1984). Also, the orbit is observed to be somewhat eccentric. Given the great age of the system, 5.1 Gy, this result is somewhat surprising. The tidal synchronization and orbital circularization theories of Zahn (1977) and Tassoul (1988) were used to estimate these timescales. The "turbulent viscosity" theory of Zahn predicts that the components of V501 Her should not have a circular orbit (timescale of 97 Gy), which is consistent with observations, but should have synchronous rotations, which is also consistent with observations (timescale of 57 My). The more efficient "hydrodynamical mechanism" of Tassoul (1988) also predicts sychronous rotations, but incorrectly predicts that the orbit should be circular (timescale of 700 Ky).

Figure 6 compares the predictions of the stellar evolution models of Yi et al. (2001) for a metallicity of $Z = 0.035$ ([Fe/H] = 0.31) and [$\alpha$/Fe] = 0.0. Evolutionary tracks are shown for the exact masses of the stars we determined. A good fit is found for an age of 5.1 Gyr. This system is unusual in that the more massive star is just beyond the core hydrogen-burning phase, while the less massive star is very near the end of core hydrogen-burning.

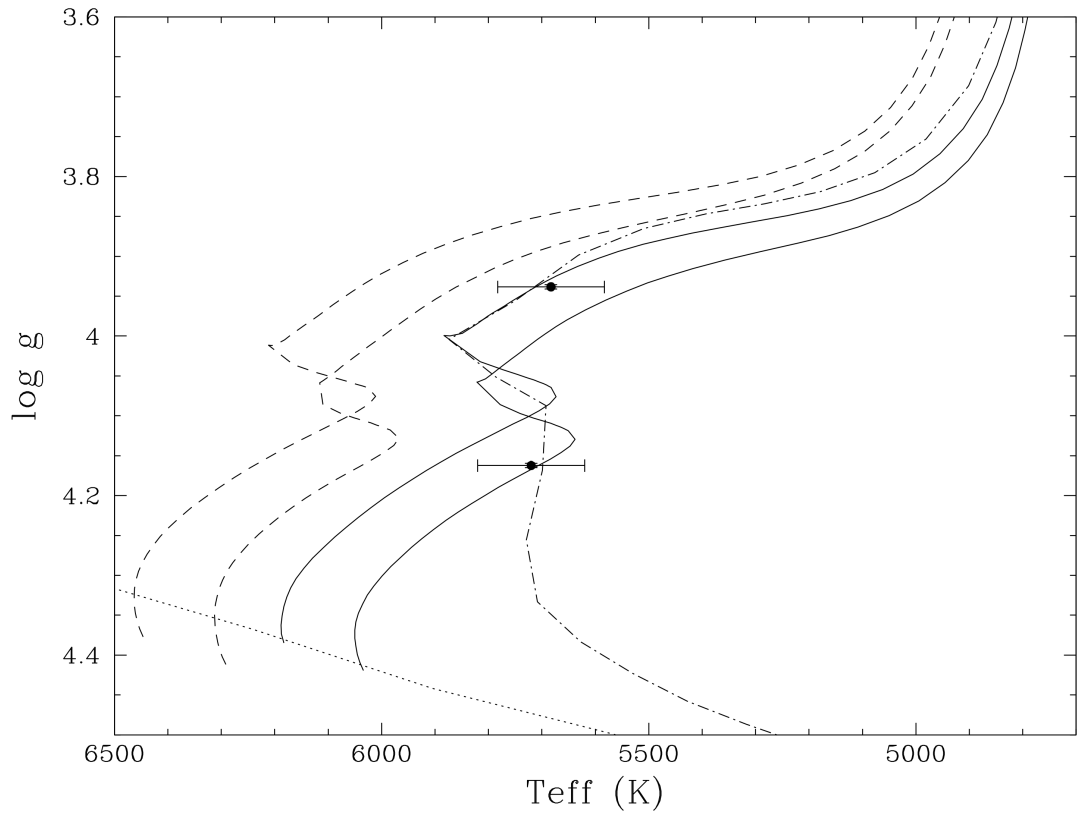

Figure 6 - Evolutionary tracks from Yi et al. (2001) compared to the observed properties of V501 Her. The two solid curves are the evolutionaly tracks for the exact masses of the stars at a metallicity of Z = 0.035. These begin at the ZAMS for that metallicity, shown as a dotted line across the lower left corner of the plot. The 5.1 Gy isochrone for that metallicity is shown as the dot-dash curve passing close to the two observed points with error bars. Corresponding evolutionary tracks for solar composition are shown as the dashed curves to the left of the best-fitting solid evolutionary tracks.

V501 Her is somewhat similar to the system V432 Aur (Siviero et al. 2004). However because of the greater difference in mass between the components of the V432 Aur system, the components of V432 Aur are farther apart in their evolution than those of

V501 Her, and so the more massive component of V432 Aur is rather more evolved than that of V501 Her. We also note that V432 Aur is a metal poor rather than a metal rich system.

ACKNOWLEDGEMENTS

The authors wish to thank Dr. A. William Neely who operates and maintains the NFO WebScope for the Consortium, and who handles preliminary processing of the images and their distribution. Astronomy at Tennessee State University is supported by the state of Tennessee through its Centers of Excellence programs. Many thanks to Guillermo Torres for providing some of the plots. Thanks also to University of Arkansas undergraduate student Craig Heinrich for initial analysis of the URSA photometry and preliminary radial velocities.